\input ./style/arxiv-general.cfg
\documentclass[aoas,MSNbibl,nameyear,seceqn,rotating,dvips]{arximspdf}
\makeatletter
   \@ifpackageloaded{graphicx}{}{\usepackage{graphicx}}
\makeatother
\usepackage{multirow}

\doi{10.1214/15-AOAS815}
\volume{9}
\issue{2}
\pubyear{2015}
\firstpage{901}
\lastpage{925}
\docsubty{FLA}

\makeatletter
\renewcommand{\mid}{|}
\makeatother

\begin{document}
\begin{frontmatter}

\title{hmmSeq: A hidden Markov model for detecting differentially expressed genes from RNA-seq data}
\runtitle{hmmSeq: A hidden Markov model}

\begin{aug}
\author[A]{\fnms{Shiqi}~\snm{Cui}\corref{}\thanksref{M1}\ead[label=e1]{scrr2@mail.missouri.edu}},
\author[A]{\fnms{Subharup}~\snm{Guha}\thanksref{M1,T2}\ead[label=e2]{guhasu@missouri.edu}},
\author[B]{\fnms{Marco A.~R.}~\snm{Ferreira}\thanksref{M2,T4}\ead[label=e3]{marf@vt.edu}} 
\and
\author[C]{\fnms{Allison N.}~\snm{Tegge}\thanksref{M2}\ead[label=e4]{ategge@vt.edu}}
\runauthor{Cui, Guha, Ferreira and Tegge}
\affiliation{University of Missouri\thanksmark{M1} and Virginia
Tech\thanksmark{M2}}
\address[A]{S. Cui\\
S. Guha\\
Department of Statistics\\
University of Missouri\\
146 Middlebush Hall\\
Columbia, Missouri 65211-6100\\
USA\\
\printead{e1}\\
\phantom{E-mail: }\printead*{e2}}
\address[B]{M.~A.~R. Ferreira\\
Department of Statistics\\
Virginia Tech\\
Blacksburg, Virginia 24061\\
USA\\
\printead{e3}}
\address[C]{A. N. Tegge\\
Computer Science\\
Virginia Tech\\
Blacksburg, Virginia 24061\\
USA\\
\printead{e4}}
\end{aug}
\thankstext{T2}{Supported by NSF Award DMS-09-06734 and by NIH Award P01CA134294.}
\thankstext{T4}{Supported by NSF Award DMS-09-07064.}

%
\received{\smonth{2} \syear{2014}}
%
\revised{\smonth{1} \syear{2015}}

%
\begin{abstract}
We introduce hmmSeq, a model-based hierarchical Bayesian technique
for detecting differentially expressed genes from RNA-seq data. Our
novel hmmSeq methodology uses hidden Markov models to account for
potential co-expression of neighboring genes. In addition, hmmSeq
employs an integrated approach to studies with technical or
biological replicates, automatically adjusting for any extra-Poisson
variability. Moreover, for cases when paired data are available,
hmmSeq includes a paired structure between treatments that
incoporates subject-specific effects. To perform parameter
estimation for the hmmSeq model, we develop an efficient Markov
chain Monte Carlo algorithm. Further, we develop a procedure for
detection of differentially expressed genes that automatically
controls false discovery rate. A~simulation study shows that the
hmmSeq methodology performs better than \mbox{competitors} in terms of
receiver operating characteristic curves. Finally, the analyses of
three publicly available RNA-seq data sets demonstrate the power and
flexibility of the hmmSeq methodology. An R package
implementing the hmmSeq framework will be submitted to CRAN upon
publication of the manuscript.
\end{abstract}

%
\end{frontmatter}

\section{Introduction}

{RNA-seq has revolutionized the study of gene expression.}
RNA-seq success may be attributed to its low noise, high-throughput
and ability to interrogate allele-specific expression and isoforms
[\citet{zhao2014comparison,auer2011differential}]. {Most} RNA-seq
studies aim to
identify 
differentially expressed (DE) genes between samples corresponding to
different treatments or biological conditions, for example, cancer
tissue versus normal tissue, {genetically engineered animals}
versus control animals, or patients exposed to two or more kinds of
treatments. These differentially expressed genes usually form the
starting point of more extensive studies such as integration of
expression data with transcription factor binding
[\citet{karlebach2008modelling}], RNA interference
[\citet{pe2011principles}] and DNA methylation
[\citet{louhimo2011cnamet}], all of which can lead to a better
understanding of regulatory mechanisms. Currently available methods
for RNA-seq data analysis assume that differential expression of genes
occurs independent of the genomic loci of each gene
[\citet{auer2011two},
\citeauthor{robinson2007moderated} (\citeyear{robinson2007moderated,robinson2008small}),
\citet{hardcastle2010bayseq},
\citet{robinson2010edger},
\citet{si2013optimal}].
However, the literature contains evidence that neighboring genes on
the chromosome tend to be
co-expressed~[\citet
{caron2001human,singer2005clusters,michalak2008coexpression}].
To account for and take advantage of this potential co-expression,
here we introduce hmmSeq.

Our hmmSeq framework incorporates potential co-expression by modeling
differential expression across the genome using hidden Markov models
(HMM). In the hmmSeq framework that we propose, neighboring gene
co-expression may occur in two ways: in differential expression across
treatments and in mean expression magnitude. Thus, we model gene
differential expression across treatments using an HMM with three
states: not differentially expressed, under- or over-expressed. This
HMM takes advantage of the potential co-differential expression by
borrowing information across neighboring genes on the chromosome. In
addition, we model gene mean expression magnitude with an HMM with two
states: low expression and high expression. The latter HMM borrows
information across the genome to increase estimation precision of the
mean expression magnitude of each gene. As we show in the simulation
study in Section~\ref{Ssimulation}, the use of information both from
neighboring genes and across the genome increases detection power and
reduces false discovery.

The existing methods for RNA-seq data analysis do not account for the
potential co-expression of neighboring
genes. \citeauthor{robinson2007moderated} (\citeyear{robinson2007moderated,robinson2008small}) use the
negative binomial distribution to model over-dispersed data through
dispersion parameters. Specifically, \citet{robinson2008small} assume
a common dispersion parameter across all tags (or genes), whereas
\citet{robinson2007moderated} assume tag-wise (or gene-wise)
dispersion parameters. To estimate these dispersion parameters, they
assume a Gaussian hierarchical hyperprior that is estimated using
empirical Bayes. After that, the gene-wise dispersion parameters are
estimated by maximum weighted likelihood. This method is implemented
in edgeR [\citet{robinson2010edger}]. The \textit{baySeq} method of
\citet{hardcastle2010bayseq} is an empirical Bayes approach that is
also based on the negative binomial distribution. An empirically
determined prior distribution is derived from the entire data set, and
rather than producing significance values, this method calculates
posterior probabilities of multiple models of differential expression,
ranking the genes by the model probabilities. \citet{blekhman2010sex}
analyze RNA-seq data by a Poisson generalized linear mixed-effect
model, which explains inter-individual variability through the
inclusion of a random individual-specific effect. Data are fitted
under the null and alternative models gene by gene, then a likelihood
ratio test is conducted to compute $p$-values, and the false discovery
rate (FDR; defined as the proportion of incorrect calls among the
genes declared as DE) is controlled by the method of
\citet{storey2003statistical}. \citet{auer2011two} have
proposed the
\textit{two-stage Poisson model} (TSPM) which assumes data contain
both overdispersed and non\-over\-dispersed genes. This technique
seeks to reduce FDR by first separating the overdispersed genes from
the non\-over\-dispersed genes, and then fitting separate models to
compute the $p$-values. \citet{benjamini1995controlling} FDR
controlling is applied on each set of $p$-values to identify DE
genes. \citet{si2013optimal} developed a test for the hypothesis that
the log fold change belongs to a subset of the real line. By assuming
parameters under a null and alternative hypothesis come from different
distributions, they estimate this mixture distribution from the data,
then the test statistic is obtained as the ratio of unconditional
probability from null parameter space over unconditional probability
from full space. All these previous methods assume that the genes are
conditionally independent. However, the exploratory data analysis we
present in Section~\ref{SEDA} suggests dependence among neighboring
genes. Our hmmSeq method addresses this dependence.

Our hmmSeq framework may also accommodate the case when there is no
dependence among the expression of neighboring genes. Specifically,
HMMs include as particular cases mixture models. In particular, the
number of components of the mixture model will be the same as the
number of states in the HMM. Thus, when there is no co-differential
expression, the result will be a mixture model with three components
that correspond to a gene being not differentially expressed, under-
or over-expressed. Likewise, when there is no dependence in mean
expression magnitude among neighboring genes, the resulting mixture
model will have two components, one component for low expression genes
and another for high expression genes. Note that the
proportion of genes in each component and the parameters of the
generating model for each component will be estimated from the
data. Thus, even without neighboring genes dependence, the hmmSeq
framework will still borrow information across the genome to learn
about each of the mixture components and, as a result, increase
estimation precision and detection power.\looseness=-1

We model extra-Poisson variability in an indirect manner. If the
experiment contains technical replicates (i.e., samples from the
same subject), then the literature provides evidence that the RNA-seq
counts are Poisson distributed [\citet
{marioni2008rna,bullard2010evaluation}]. On the other hand, if the experiment
contains biological replicates, then the RNA-seq counts will have
extra-Poisson variability [\citet
{langmead2010cloud,robinson2007moderated}]. This extra-Poisson
variability may be a
result of across-subjects variability or slight differences in the
experimental conditions when the samples were taken or analyzed. While
the RNA-seq literature usually uses the negative binomial distribution
to model the extra-Poisson variability, another way to deal with this
extra-variability is through the use of the Poisson distribution
together with random effects. We prefer the latter because it provides
a framework that can flexibly deal with known sources of extra-Poisson
variability such as, for example, biological variation among subjects.
In the case of paired data
considered in Section~\ref{Spairedreplicate}, we deal with the
biological variation
by including for each gene subject-specific random effects. Moreover,
for nonpaired
data we implicitly deal with the subject-specific random effects (and
any other source of extra-Poisson variability) by assuming that for
nondifferentially expressed genes the differential treatment effect
parameter may come from a normal distribution centered at
zero. Hence, for nondifferentially expressed genes the differential
treatment effect parameter is a sum of the random effects of subjects
and other hidden sources. In addition to facilitating the
implementation of our HMM framework, this aspect of our model
increases robustness with respect to hidden unforeseen sources of
variability.

We have investigated in three fronts the practical usefulness and
adequacy of HMMs and mixture components models for RNA-seq data
analysis. First, we have performed an exploratory data analysis
presented in Section~\ref{SEDA} that studies for two real RNA-seq
data sets the empirical statistical properties of preliminary estimates
of differential expression parameters and mean expression magnitude
parameters. Two patterns emerge from this exploratory data analysis:
the existence of three differential expression states and of two mean
expression magnitude states; and a possible dependence across
neighboring genes. Second, we have performed a simulation study that
considers all four possible combinations of HMMs and mixture
components models for differential expression and mean expression
magnitude. This simulation study compares the performance of our
hmmSeq framework with competing RNA-seq analysis methodologies. In all
four possible cases, our hmmSeq framework beats the competing methods
in terms of receiver operating characteristic curves. Finally, we
have used the deviance information criterion (DIC)
[\citet{spiegelhalter2002bayesian}] to decide among the four possible
combinations of HMMs and mixture components models what is the most
adequate model for each of three real RNA-seq data sets. Our use of the
DIC is justified by its good performance in a simulation study
presented in Section~\ref{Ssimulation}. The DIC indicates dependence
across neighboring genes for two of the three data sets. Therefore, in
this paper we provide further evidence that for some biological
processes neighboring genes on the chromosome tend to be co-expressed.

We take a full Bayesian analysis approach and develop a Markov chain
Monte Carlo algorithm to exploit the posterior distribution of the
model parameters. To simulate the differential effects and the mean
effect magnitudes, we develop an efficient Metropolis--Hastings
algorithm for hidden Markov models. In addition, we use the output of
the MCMC algorithm to identify differentially expressed genes while
controlling for false discovery rate. Specifically, we use a Bayesian
approach for controlling the FDR level proposed by
\citet{newton2004detecting} and further studied by
\citet{muller2007fdr}. We demonstrate the advantages and benefits of
our hmmSeq methodology by analyzing three RNA-seq data sets. The first
data set [\citet{marioni2008rna}] consists of five technical replicates
each of a kidney and liver RNA sample. The second data set
[\citet{zeng2012divergent}] consists of six biological replicates
extracted from two regions, frontal pole and hippocampus, of normal
human brains. Finally, the third data set [\citet{henn2013high}] consists
of paired B-cell samples data of day 0 (before vaccination) and day 7
(\mbox{post-}vaccination) for 3 pre-vaccinated subjects. Therefore, we
demonstrate the power and flexibility of the hmmSeq methodology on
three types of RNA-seq data: technical replicates, biological
replicates and paired samples.\looseness=-1



The remainder of the paper is organized as follows. Section~\ref
{Smodel} describes the details of the hmmSeq model and informally
demonstrates the necessity for hidden Markov models in Section~\ref
{SEDA}. Section~\ref{Sinference} describes the posterior
inference procedure based on Markov chain Monte Carlo (MCMC)
techniques and the procedure for identification of DE genes.
Section~\ref{Ssimulation} uses simulated data to demonstrate the
effectiveness of hmmSeq relative to well-known techniques for RNA-seq.
The \citet{marioni2008rna}, \citet{zeng2012divergent} and
\citet{henn2013high} data sets are analyzed in Sections~\ref
{Stechreplicate},~\ref{Sbioreplicate}~and~\ref{Spairedreplicate}. In
all cases, the results are
compared and contrasted with those of existing approaches to
demonstrate the success of hmmSeq. A functional analysis of the
detected sets of DE genes provides further evidence of the reliability
of the procedure. {An R package implementing the hmmSeq framework
will be submitted to CRAN upon publication of the manuscript}.

\section{A Bayesian hierarchical model for RNA-seq data}\label{Smodel}

We focus on two-treatment comparisons. For a given chromosome $c$, let
$Y_{ijkc}$ denote the integer-valued gene read of the $k$th replicate
of gene $i$ under treatment $j$, for gene $i=1,2,\ldots,I_{c}$,
treatment $j=1,2$, and replicate $k=1,2,\ldots,K_{j}$ on chromosome~$c$. The genes are sequentially arranged so that consecutive indices
correspond to neighboring genes on the chromosome. We assume that
%
\begin{eqnarray}\label{lambda}
Y_{ijkc}&\stackrel{\mathrm{indep}}\sim&\operatorname{Poisson}(\lambda_{ijkc})
\qquad\mbox{where }
\nonumber
\\
\log(\lambda_{i1kc} )&=&\beta_{ic}-\Delta_{ic}+
\rho_{1k}\quad\mbox{and}
\\
\log(\lambda_{i2kc} )&=&\beta_{ic}+\Delta_{ic}+
\rho_{2k},\nonumber
\end{eqnarray}
where $\beta_{ic}$ denotes the mean expression magnitude of gene $i$
and $2\Delta_{ic}$ denotes the $\log$-fold change between the
treatments. The treatment-specific replicate effects are represented
by $\rho_{jk}$. We observe that the treatments are a priori
interchangeable in equation~(\ref{lambda}). The \textit{differential
treatment effect} $\Delta_{ic}$ for gene $i$ is key because it
determines the relative expression levels of the treatments for the
gene. That is, $\Delta_{ic}$ determines whether treatment 2 is over-,
under-, or nondifferentially expressed with respect to treatment 1.

To model the mean expression magnitude, the possible dependence among
the $\beta_{ic}$'s of neighboring genes on a chromosome is modeled
using either a two-component finite mixture model
[\citet{titterington1985statistical,fruhwirth2006finite}] or a
stationary two-state hidden Markov model [\citet
{rabiner1989tutorial,macdonald1997hidden}]. The latent \textit{average
expression state}
$s_{ic}$ determines whether the expression of gene $i$, averaged over
treatments and replicates, is ``small'' ($s_{ic}=1$) or ``large''
($s_{ic}=2$). In the absence of differential treatment and replicate
effects, the two levels of this categorical variable correspond,
respectively, to low and high reads for the genes. Conditional on
$s_{ic}$, the average expression $\beta_{ic}$ is normally distributed:
\[
\beta_{ic}|s_{ic} \stackrel{\mathrm{indep}} \sim N\bigl(
\mu_{s_{ic}c},\sigma_{s_{ic}c}^2\bigr) 
\]
with $\mu_{1c} < \mu_{2c}$. The latent states
$s_{1c},\ldots,s_{I_{c}c}$ follow either a finite mixture model (FMM)
with probability vector $\mathbf{P}_{c}=(p_{1c},p_{2c})$ or a
hidden Markov model (HMM) with stationary transition probability
matrix $\mathbf{A}_{c}=((a_{utc}))_{2\times2}$ with the row sums
$\sum_{t=1,2} a_{utc}=1$ for $u=1,2$. We denote the two-component FMM
by $\mathcal{F}_{2c}$ and the two-state HMM by $\mathcal{H}_{2c}$, assuming
independent, noninformative priors for its dispersion parameters:
$p(\sigma_{uc}^2)\propto\sigma_{uc}^{-2}$ for $u=1,2$.

To model differential expression, the differential effects
$\Delta_{1c},\ldots,\Delta_{I_{c}c}$ of the genes are modeled either
by a finite mixture model (FMM) $\mathcal{F}_{3c}$ with probability
vector $\mathbf{Q}_{c}=(q_{1c},q_{2c},q_{3c})$ or by a
three-state stationary HMM denoted by $\mathcal{H}_{3c}$; the matrix
of transition probabilities is denoted by
$\mathbf{B}_{c}=((b_{vtc}))_{3\times3}$ with row sums $\sum_{t=1}^3
b_{vtc}=1$ for $v=1,2,3$. With latent \textit{differential states}
$h_{1c},\ldots,h_{I_{c}c}$ taking values in $\{1,2,3\}$, the values
correspond, respectively, to the gene-specific under-,
\mbox{nondifferential-,} and over-expression of treatment 2 relative
to treatment 1. Given the state $h_{ic}$, differential effect
$\Delta_{ic}$ is distributed as
%
\begin{equation}
\label{eqnhmm-DE}
\qquad\Delta_{ic}\mid h_{ic}\sim\cases{ N\bigl(
\phi_{1c},\tau_{1c}^2\bigr),&\quad if
$h_{ic}=1$ (under-expressed),
\vspace*{3pt}\cr
N\bigl(0,\tau_{2c}^2
\bigr), &\quad if $h_{ic}=2$ (nondifferentially-expressed),
\vspace*{3pt}\cr
N\bigl(
\phi_{3c},\tau_{3c}^2\bigr), &\quad if
$h_{ic}=3$ (over-expressed),}
\end{equation}
where $\phi_{1c}<0$ and $\phi_{3c}>0$. Thus, for each chromosome,
$h_{1c},\ldots,h_{I_{c}c}$ are the parameters of interest because they
identify the set of DE genes.

We observe that the latent states of both FMM $\mathcal{F}_{2c}$ and
HMM $\mathcal{H}_{2c}$ are nonexchangeable, being associated with
particular biological conditions. The priors for the state parameters
are designed to reflect this and also to prevent label switching
[\citet{scott2002bayesian}]. Specifically, the mean parameters,
$\mu_{1c}$ and $\mu_{2c}$ are assigned the prior
$p(\mu_{1c},\mu_{2c})\propto1_{\{\mu_{1c} \le\mu_{2c}-\delta\}}$
where $\delta>0$ is a predetermined constant. The fact that $\delta$
is strictly positive guarantees that $\mu_{1c} < \mu_{2c}$ and the two
states are identifiable.

For the same reason, for the FMM $\mathcal{F}_{3c}$ and HMM
$\mathcal{H}_{3c}$, we assume that $p(\phi_{1c},\phi_{3c})\propto
1_{\{\phi_{1c}<u_1, \phi_{3c}>l_3\}}$, where $u_1<0$ and $l_3>0$ are
prespecified constants that can be chosen as follows. The $\log$-fold
change between the over- and under-expressed categories is at least
$(l_3-u_1)$. From a practical standpoint, in order to distinguish
between these two categories, it is reasonable to assume that the
ratios of their associated $\Delta_{ic}$'s exceed $2$. Because of
this, we symmetrically set $u_1=-(\log{2})/2$ and $l_3=(\log{2})/2$.
To further facilitate inferences of the state-specific parameters,
informative conjugate priors are assigned to $\tau_{1c}^2$,
$\tau_{2c}^2$ and~$\tau_{3c}^2$.

\subsection{Paired data analysis}\label{Spaired}

Our hmmSeq framework may also accommodate paired data, that is, the
case when each subject undergoes each of the treatments. Here we
describe the minor changes needed for that purpose. For a given
chromosome $c$, let $Y_{ijkc}$ denote the gene read of the $k$th
subject of gene $i$ under treatment~$j$, for subject
$k=1,\ldots,K$. Obviously, because of the paired data structure there
exists dependence between observations on the same subject. To account
for this dependence, we assume that
\begin{eqnarray}\label{lambdapaired}
Y_{ijkc} & \stackrel{\mathrm{indep}} \sim& \operatorname{Poisson}(\lambda_{ijkc})
\qquad\mbox{where }
\nonumber
\\
\log(\lambda_{i1kc} )&=&\beta_{ic}-\Delta_{ic}+
\varepsilon_{ikc}+\rho_{1k}\quad\mbox{and }
\\
\log(\lambda_{i2kc} )&=&\beta_{ic}+\Delta_{ic}+
\varepsilon_{ikc}+\rho_{2k},\nonumber
\end{eqnarray}
with $\varepsilon_{ikc}\sim N(0,\sigma_{\varepsilon}^2)$ denoting the
subject-specific random effects. The other parameters in the
paired-data model have the same interpretations as in
equation~(\ref{lambda}).

\subsection{Exploratory data analysis}\label{SEDA}

We have performed an exploratory data analysis (EDA) to verify some of
the hmmSeq model assumptions for the \citet{marioni2008rna} and
\citet{zeng2012divergent} data sets. Both data sets possess the feature
described by \citet{bullard2010evaluation};
for 
libraries under each treatment, 5\% of the genes account for over 50\%
of the total library size, and 10\% of the genes account for over 60\%
of the total library size. Focusing only on those genes whose reads
exceeded nine for all libraries, and ignoring the replicate effects,
we preprocessed the raw counts using the {upper-quartile}
normalizing technique of \citet{bullard2010evaluation}.

%
\begin{figure}[t]

\includegraphics{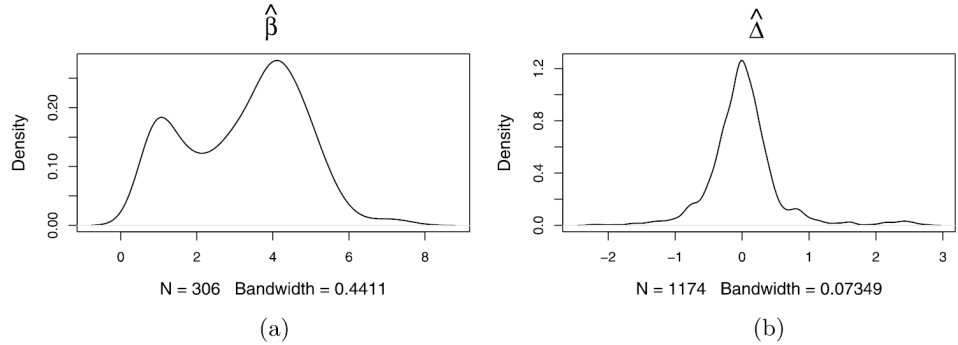}

\caption{\citet{marioni2008rna} data set---density plots of
preliminary estimates for the expression magnitudes $\beta$ and
the differential expression effects $\Delta$.
\textup{(a)}~Density plot of $\hat{\beta}_i$ for chromosome 13,
\textup{(b)}~density plot of $\hat{\Delta}_i$ for chromosome 19.}\label{EDAtech}\label{TBETADEN}\label{TDELTADEN}
\end{figure}

For the genes $i=1,2,\ldots,I_{c}$ of each chromosome, we have
computed preliminary estimates of the expression magnitude
$\beta_{ic}$ and differential effect $\Delta_{ic}$ by treating these
parameters as the fixed effects in a Poisson regression~model.
Figure~\ref{EDAtech} displays graphical summaries of these estimates
for a
few chromosomes of the \citet{marioni2008rna} data set. The results
were similar for the other chromosomes. The density plot for the
$\beta_i$'s in Figure~\ref{TBETADEN}(a) and for the $\Delta_i$'s in
Figure~\ref{TDELTADEN}(b) are indicative of mixture of densities
representations for these parameters. Further, data analysis in
Section~\ref{Stechreplicate} will select the dependence structures
of $\beta_{i}$'s and $\Delta_{i}$'s by DIC model selection.


%
\begin{figure}[b]

\includegraphics{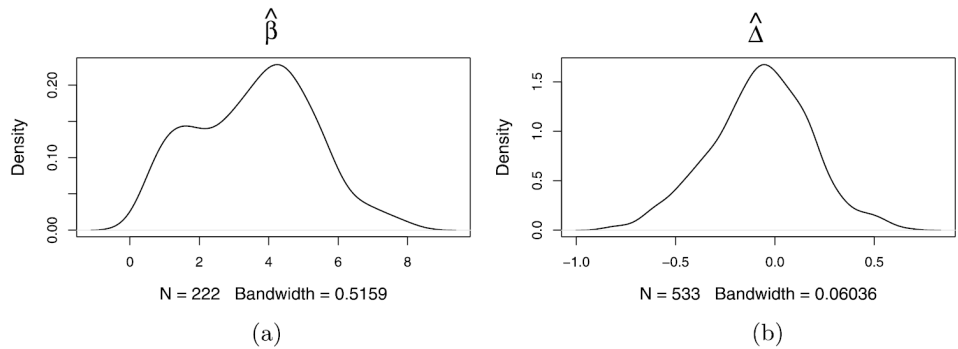}

\caption{\citet{zeng2012divergent} data set---density plots of
preliminary estimates for the expression magnitudes $\beta$ and
the differential expression effects $\Delta$.
\textup{(a)}~Density plot of $\hat{\beta}_i$ for chromosome 13,
\textup{(b)}~density plot of $\hat{\Delta}_i$ for chromosome 9.}\label{EDAbio}
\end{figure}

For the \citet{zeng2012divergent} data set, Figure~\ref{EDAbio}
displays density plots for preliminary estimates of the model
parameters and also reveals a similar~pattern as the
\citet{marioni2008rna} data set. The above analysis, together with
data analysis in Section~\ref{Sbioreplicate}, suggests the need for
mixture models, with first order dependence to model the expression
magnitudes $\beta_{ic}$ and differential effects $\Delta_{ic}$,
justifying the use of the hidden Markov models $\mathcal{H}_{2c}$ and
$\mathcal{H}_{3c}$ in the hmmSeq method.

\section{Posterior inference}\label{Sinference}

We investigate the posterior distribution of the chromosome-specific
parameters using Markov chain Monte Carlo (MCMC) methods. Gibbs
sampling cannot be applied to generate the parameters in equation
(\ref{lambda}) because the Poisson likelihood function is not
conjugate to the normal priors of the parameters. Consequently, we
apply the Laplace approximation [e.g., \citet
{zeger1991generalized,chib1994bayes}] to generate proposed updates for
the equation
(\ref{lambda}) parameters. The proposals are accepted or rejected by a
Metropolis--Hastings probability to compensate for the use of an
approximation instead of the Poisson distribution. This guarantees the
convergence of the Markov chain to the posterior distribution of the
hmmSeq model. We analyze each chromosome separately and, for
simplicity of notation, in this section we omit the chromosome index
$c$.

\subsection{Metropolis--Hastings algorithm for HMMs}\label{secmetr-hast-algor}

In this section we present a Metropolis--Hastings algorithm for the
simulation of a general latent process $\{\theta_i, i = 1, \ldots,
I\}$ that follows an HMM $\mathcal{H}_m$. We use this algorithm in
Section~\ref{secmcmc-procedure} to simulate the expression magnitude
$ \bolds{\beta}$ and the differential effects
$\bolds{\Delta}$. Under a Laplace approximation, the
\textit{working values} of the read counts are defined as
%
\begin{eqnarray}
w_{ijk} &=& \log(\lambda_{ijk} ) + \frac{Y_{ijk}-\lambda_{ijk}}{\lambda_{ijk}}.
\label{laplace}
\end{eqnarray}
These working values have an approximate normal distribution,
specifically $ w_{ijk} \stackrel{\mathrm{approx}}\sim
N (\log(\lambda_{ijk}), 1/\lambda_{ijk} )$.

For a more general case, suppose $\theta_i$ is the parameter of
interest, and its value at the previous MCMC iteration was
$\theta_i^{(\mathrm{old})}$. Then, the Laplace approximation (\ref{laplace})
gives us $w_{ijk}=\log{(\lambda_{ijk}^{*})}+
(Y_{ijk}-\lambda_{ijk}^*)/\lambda_{ijk}^*$
$\stackrel{\mathrm{approx}}\sim
N(\log (\lambda_{ijk}),\break 1/\lambda_{ijk}^*)$, where
$\log{(\lambda_{ijk})} = \xi_{ijk}+z_{ijk}\cdot\theta_i$ and
$\log{(\lambda_{ijk}^*)} = \xi_{ijk}+z_{ijk}\cdot\theta_i^{(\mathrm{old})}$.

Let the vectors
$\mathbf{w}_i=(w_{i11},w_{i12},\ldots,w_{i2K_{2}})^{\prime}$,
$\bolds{\lambda}_i=(\lambda_{i11},\lambda_{i12},\ldots,\lambda
_{i2K_{2}})^{\prime}$
and
$\bolds{\lambda}_i^{*}=(\lambda_{i11}^{*},\lambda_{i12}^{*},\ldots
,\lambda_{i2K_{2}}^{*})^{\prime}$. Then
$\mathbf{w}_i\stackrel{\mathrm{approx}}\sim
N(\log{(\bolds{\lambda}_{i})},\operatorname{Diag}(1/\bolds{\lambda}_i^{*}))$.\break
Defining
\[
w_i^*=\frac{\sum_{j=1}^{2}\sum_{k=1}^{K_{j}}\lambda
_{ijk}^*[z_{ijk}(w_{ijk}-\xi_{ijk})]}{\sum_{j=1}^{2}\sum
_{k=1}^{K_{j}}z_{ijk}^2\lambda_{ijk}^*},
\]
we\vspace*{2pt} have that $w_i^*$ is sufficient for $\theta_i$ and $w_i^*\mid
\theta_i\stackrel{\mathrm{approx}}\sim
N(\theta_i,1/\sum_{j=1}^{2}\sum_{k=1}^{K_{j}}z_{ijk}^2\lambda_{ijk}^*)$.
Further assume that the prior of $\delta$ is an $m$-state hidden
Markov model ($\mathcal{H}_m$) with transition matrix
$\mathbf{C}_m$, and $\theta_i\mid h_i=t\sim N(\nu_t,\kappa_t^2)$,
for $t=1,2,\ldots,m$, where $h_i$ is the hidden state for
$\theta_i$. We marginalize over $\theta_i$ to obtain the approximate
likelihood function
%
\begin{eqnarray}
p\bigl(w_i^*|h_i=t\bigr)\stackrel{\mathrm{approx}}\sim
N\Biggl(\nu_t,\kappa_t^2+1\Big/\sum
_{j=1}^{2}\sum_{k=1}^{K_{j}}z_{ijk}^2
\lambda_{ijk}^*\Biggr)
\nonumber\\[-9pt]\\[-9pt]
\eqntext{\mbox{where } t=1,2,\ldots,m.}\label{wi^*}
\end{eqnarray}
The conditional prior probability, $P(h_i=t | h_j, j \neq i)$, for
$t=1,2,3$, can be computed from the transition matrix,
$\mathbf{C}_m$, of the HMM $\mathcal{H}_m$.

The normalized product of the conditional prior probability and
approximation~(\ref{wi^*}) gives the approximate full conditional
distribution of the differential state $h_i$, from which we propose a
new value, $h_i^{(\mathrm{prop})}$. We then propose a new value,
$\theta_i^{(\mathrm{prop})}$, from the approximate full conditional of
$\theta_i$ given $h_i=h_i^{(\mathrm{prop})}$. The proposed values
$(h_i^{(\mathrm{prop})},\theta_i^{(\mathrm{prop})})$ are jointly accepted
or rejected by a Metropolis--Hastings probability
[\citet{gamerman2006markov}] to ensure that the post-burn-in MCMC
samples represent draws from model posterior.

\subsection{MCMC procedure}\label{secmcmc-procedure}

We iteratively generate MCMC samples of the chromosome-specific
parameters by the following procedure:
\begin{longlist}[2.]
\item[1.] The differential effects
$\Delta_1,\ldots,\Delta_I$ and latent differential states
$h_1,\ldots,h_I$ are generated as in
Section~\ref{secmetr-hast-algor}, given the expression magnitudes
$\bolds{\beta}$, subject-specific effects~$\bolds{\varepsilon}$ (set to be $0$ for nonpaired data) and
treatment-replicate effects $\bolds{\rho}$.
\item[2.] The mean expression magnitudes
$\beta_1,\ldots,\beta_I$ and latent states $s_1,\ldots,s_I$ are
generated as in Section~\ref{secmetr-hast-algor}, given the
differential effects $\bolds{\Delta}$, subject-specific effects~$\bolds{\varepsilon}$ (set to be $0$ for nonpaired data) and
treatment-replicate effects $\bolds{\rho}$.

\item[3.] For paired data, the subject-specific effects $\varepsilon_{ik}$
for $i=1,\ldots,I$ and $k=1,\ldots,K$ are also generated by a
similar Laplace approximation and Metropolis--Hastings procedure as
in step~1.

\item[4.] Conditional on the mean expression magnitudes
$\beta_1,\ldots,\beta_I$, latent states $s_1,\ldots,s_I$, and the
fact that $\mu_2-\mu_1>\delta$, the hyperparameters $\mu_1$ and
$\mu_2$ are jointly sampled from the restricted bivariate normal
distribution using the R package \emph{tmvtnorm}
[\citet{wilhelm2013tmvtnorm}].

\item[5.] For latent states $h=1,2,3$, $s=1,2$, the hyperparameters
$\phi_h$, $\tau_h^2$, $\sigma_s^2$ are all generated from their full
conditional distributions by Gibbs sampling steps.
\end{longlist}

\subsection{Detection of DE genes}\label{Sdetect}

For each chromosome, interest focuses on the latent vector of
differential states $h_1,\ldots,h_I$, where, as defined in
equation~(\ref{eqnhmm-DE}), $h_i=1$ ($h_i=3$) represents an
under-expressed (over-expressed) gene in treatment~2. We use the MCMC
samples of the differential states to identify the DE genes while
controlling for false discovery rate. Specifically, we use a Bayesian
approach for controlling the FDR level first proposed by
\citet{newton2004detecting}, further studied by \citet{muller2007fdr},
and subsequently applied in RNA-seq analysis by
\citet{lee2011differential}.

Let $q_0$ be the desired nominal FDR level. In addition, let $r_i\in
\{0,1\}$ represent the unknown truth that gene $i$ is differentially
expressed ($r_i=1$) or nondifferentially expressed ($r_i=0$). Further,
let $p_i$ be the posterior probability that gene $i$ is differentially
expressed. Last, let $\delta_i\in\{0,1\}$ denote the decision of
calling gene $i$ differentially expressed ($\delta_i=1$) or
nondifferentially expressed ($\delta_i=0$). Using the MCMC output, we
compute the estimate $\hat{p}_i=\operatorname{Pr}(r_i=1\mid \mbox{data})$ for
genes $i=1,\ldots,N$ on all chromosomes.

A possible decision is to flag all genes with $\hat{p}_i$ greater than
or equal to a certain threshold $p_0$. The resulting FDR would then be
equal to
%
\begin{equation}
\label{eqFDR} \mbox{FDR}=\frac{\sum_{i=1}^N(1-r_i)1_{(\hat{p}_i \geq
p_0)}}{\sum_{i=1}^N
1_{(\hat{p}_i \geq p_0)}}.
\end{equation}
Hence, the posterior expected FDR would be
%
\begin{equation}
\label{eqEFDR} \widehat{\mbox{FDR}}=\frac{\sum_{i=1}^N(1-\hat
{p}_i)1_{(\hat{p}_i \geq p_0)}}{\sum_{i=1}^N
1_{(\hat{p}_i \geq p_0)}}.
\end{equation}

Alternatively and more effectively than assigning a prespecified
threshold $p_0$, we may control the nominal FDR level $q_0$
[\citet{newton2004detecting,muller2007fdr}]. Specifically, first we
rank genes in decreasing order of $\hat{p}_i$. Denote the ordered
estimated posterior probabilities by $\hat{p}_{(1)} > \hat{p}_{(2)} >
\cdots> \hat{p}_{(N)}$. Thus, if we declare as differentially
expressed the set of genes such that $\hat{p}_i \geq\hat{p}_{(d)}$,
for each $d=1,\ldots,N$, then the corresponding posterior expected FDR
will be
%
\begin{equation}
\label{eqEFDRd} \widehat{\mbox{FDR}}_d=\frac{\sum_{i=1}^N(1-\hat
{p}_i)1_{(\hat{p}_i \geq p_{(d)})}}{\sum_{i=1}^N
1_{(\hat{p}_i \geq p_{(d)})}} =
\frac{\sum_{i=1}^d(1-\hat{p}_{(i)})}{d}.
\end{equation}
Finally, the decision rule for detecting DE genes is to flag all genes
with \mbox{$\widehat{\mbox{FDR}}_d<q_0$}.

\section{Simulation study}\label{Ssimulation}

{To compare the accuracy of hmmSeq with existing RNA-seq
techniques, we performed two simulation studies. In the first study,
the data were generated from a Poisson distribution. In the second
simulation study, we generated the data from a negative binomial
distribution. Three popular RNA-seq techniques were considered for
comparisons: edgeR [\citet{robinson2010edger}], baySeq
[\citet{hardcastle2009bayseq}], and TSPM [\citet
{auer2011two}].} The
methods edgeR and baySeq have been implemented in R packages publicly
available at \surl{http://www.bioconductor.org}. R code for TSPM can be
downloaded from \surl{http://www.stat.purdue.edu/\textasciitilde
doerge/software/TSPM.R}. The R code for\break
hmmSeq is available in the Supplementary Materials [\citet{supp}].

%
\begin{table}[t]
\tabcolsep=0pt
\caption{Simulation study---parameters for generation of $\bolds{\beta}$}\label{tabparametersbeta}
\begin{tabular*}{\tablewidth}{@{\extracolsep{\fill}}lcccccc@{}}\hline
& \multicolumn{4}{c}{\textbf{Normal components}}& \textbf{HMM transition matrix} &
\textbf{FMM probability}\\
\hline
$\bolds{\beta}$& $\mu_1$ & $\sigma_1^2$ & $\mu_2$ & $\sigma_2^2$ & $\mathbf{A}$ & $P$\\
& 1 & 0.37 & 3.91 & 2.4 & $\pmatrix{ 0.50 & 0.50 \cr0.05& 0.95}$ & $(0.1,0.9)$\\
\hline
\end{tabular*}
\end{table}
%

%
\begin{table}[b]
\tabcolsep=0pt
\caption{Simulation study---parameters for generation of $\bolds{\Delta}$}\label{tabparametersdelta}
\begin{tabular*}{\tablewidth}{@{\extracolsep{\fill}}@{}lccccccc@{}}
\hline
& \multicolumn{5}{c}{\textbf{Normal components}}& \textbf{HMM transition matrix} & \textbf{FMM probability}\\
\hline
$\bolds{\Delta}$& $\phi_1$ & $\tau_1^2$ & $\tau_2^2$ & $\phi_3$ & $\tau_3^2$ & $\mathbf{B}$ & $Q$\\
& $-$0.4 & 0.013 & 0.01&0.4 & 0.013 & $\pmatrix{ 0.50 & 0.25 & 0.25 \cr
0.10& 0.80&0.10\cr0.25&0.25&0.50}$ & $(0.22,0.56,0.22)$\\
\hline
\end{tabular*}
\end{table}

{We first consider a simulation study with data generated from a
Poisson distribution.} For each of the following simulations, read
counts were simulated for 12 chromosomes having 800 genes each,
resulting in a total of 9600 genes. Six replicates of the set of
read counts were generated for each of the two treatments. The
replicate effects were assumed to be equal to the estimates for the
biological replicates data of \citet{zeng2012divergent}.

In the simulation study, the gene-specific magnitude factors
$\bolds{\beta}$ and the differential expression factors
$\bolds{\Delta}$ were generated either from the hidden Markov model
or finite mixture model with hyperparameters values given in
Tables~\ref{tabparametersbeta} and \ref{tabparametersdelta}.
{The hyperparameters of the normal components were chosen to match
the estimates for the biological replicates data of
\citet{zeng2012divergent}. The other hyperparameters were chosen
according to our experience working with
hidden Markov models [\citet{guha2008bayesian}].}
Let ``F'' denote a finite mixture model and ``H'' denote a hidden Markov
model. We consider each model resulting from each possible combination
of a FMM or a HMM for $\bolds{\beta}$ and a FMM or a HMM for
$\bolds{\Delta}$ in a total of 4 possible models. We denote each
model with FF, FH, HF and HH, with the first letter indicating the
process for $\bolds{\beta}$ and the second letter indicating the
process for $\bolds{\Delta}$.

We perform model selection with the deviance information criterion
(DIC) [\citet{spiegelhalter2002bayesian}]. To evaluate the ability of
DIC to discriminate among the four competing models, we have performed
a simulation study. Specifically, for each of the 4 possible models,
we have simulated 30 data sets. After that, we have analyzed each
simulated data set with the 4 different hmmSeq models: FF, FH, HF and
HH. Then, for each simulated data set we have conducted DIC model
selection. Table~\ref{tabDIC} presents for each true model the
proportion of times that DIC has chosen each of the 4 competing
models. As we can see from Table~\ref{tabDIC}, the DIC chooses the
correct model most of the time.

%
\begin{table}[t]
\tabcolsep=10pt
\caption{Simulation study---performance of DIC-based model selection}\label{tabDIC}
\begin{tabular*}{\tablewidth}{@{\extracolsep{\fill}}@{}lccccc@{}}
\hline
& & \multicolumn{4}{c@{}}{\textbf{DIC chosen model}}\\[-6pt]
& & \multicolumn{4}{c@{}}{\hrulefill}\\
& & \textbf{FF} & \textbf{FH} & \textbf{HF} & \textbf{HH}\\
\hline
True model& FF & 0.73 & 0.10 & 0.07 & 0.10\\
& FH & 0.00 & 0.77 & 0.00 & 0.23\\
& HF & 0.07 & 0.03 & 0.80 & 0.10\\
& HH & 0.00 & 0.10 & 0.00 & 0.90\\
\hline
\end{tabular*}
\end{table}

\begin{figure}

\includegraphics{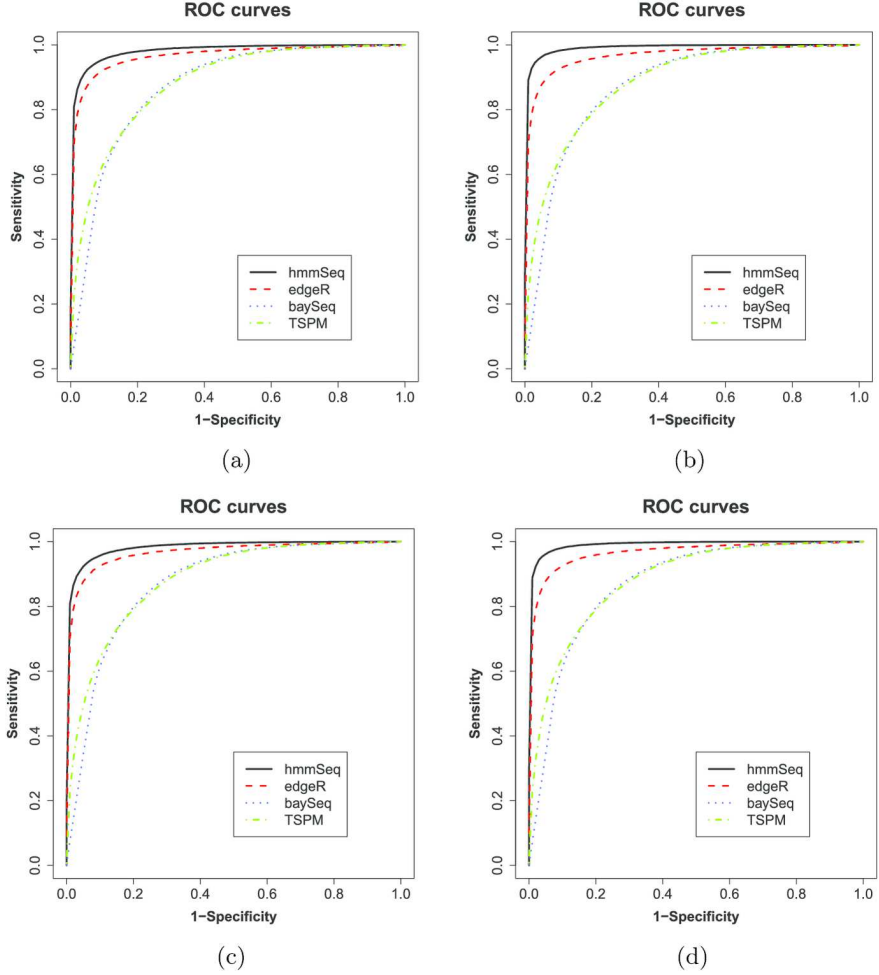}

\caption{For the simulation study of four setups FF~\textup{(a)}, FH~\textup{(b)}, HF~\textup{(c)} and
HH~\textup{(d)}, four panels depict the ROC curves for hmmSeq, edgeR, baySeq
and TSPM. Results are averaged over 30 simulated data sets for
each setup, where for each simulated data set we use the
DIC-chosen hmmSeq model.}\vspace*{-1pt}\label{ROC}
\end{figure}

To compare hmmSeq with the other competing RNA-seq analysis methods,
we consider their receiver operating characteristic (ROC) curves. The
ROC curve of each method describes the relationship between the true
positive rate (TPR) and~the false positive rate (FPR) of gene
detection. The TPR (which is also known as the sensitivity) is defined
as the proportion of truly DE genes that are detected by the
method. The FPR is defined as the proportion of non-DE genes that are
erroneously identified as DE. The greater the area under the ROC
curve, the greater the reliability of the method in detecting
DE~genes. For each simulation setup of FF, FH, HF and HH, we plot the
ROC curves of (DIC picked) hmmSeq, edgeR, baySeq and TSPM averaged
over 30 repetitions. Figure~\ref{ROC} displays the ROC curves for the
methods hmmSeq (solid line), edgeR (dashed line), baySeq (dotted
line) and TSPM (dot-dashed line) with the areas below the ROC curves
being indicative of the relative accuracies of the methods in
detecting DE genes. While edgeR beat the methods TSPM and baySeq in
this simulation, hmmSeq achieves a substantially higher area under the
ROC curve than the competing methods.

%

%
\begin{figure}

\includegraphics{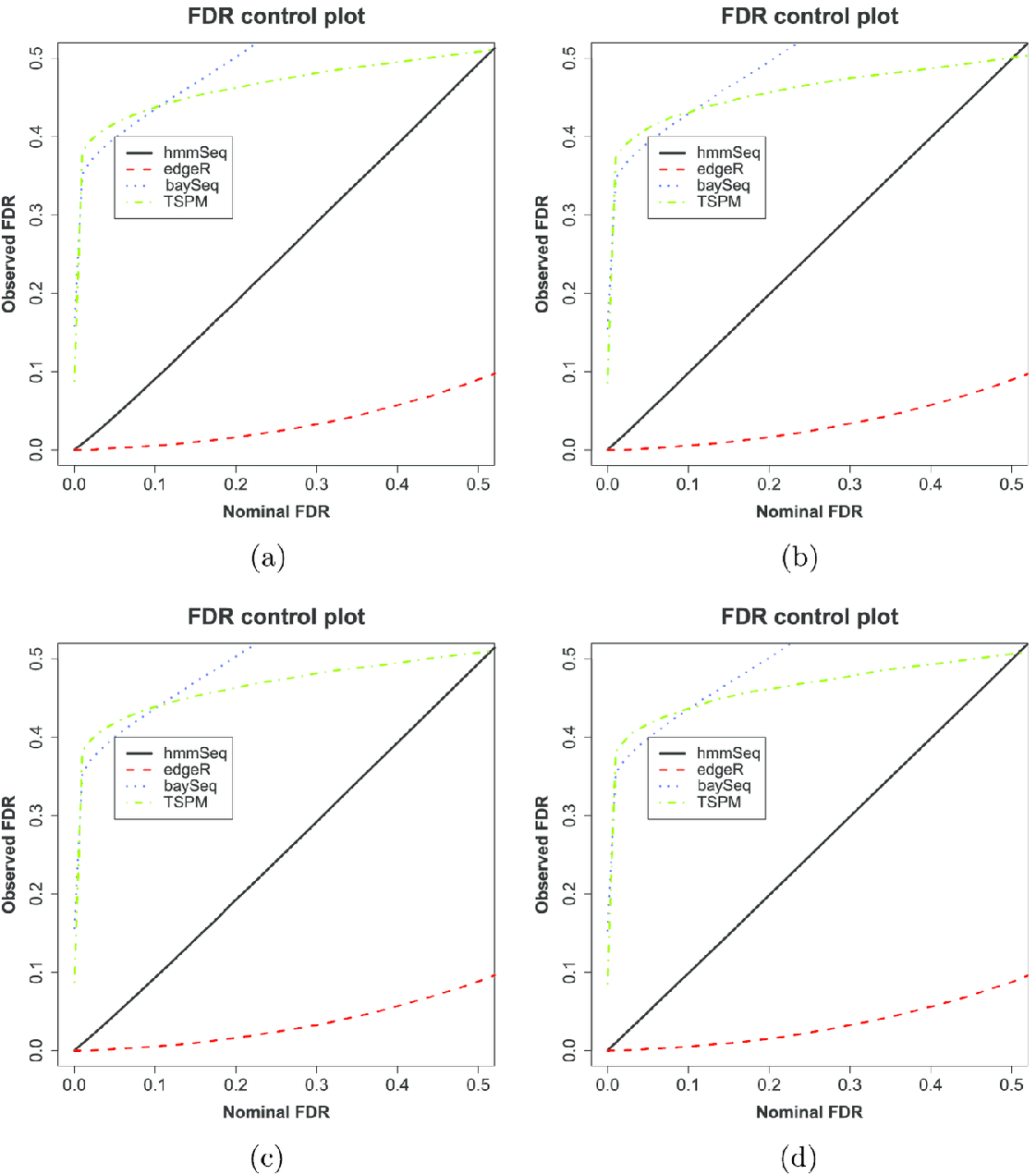}

\caption{For the simulation study, four panels depict the observed
FDR versus nominal FDR for the methods hmmSeq, TSPM, edgeR and
baySeq under four different simulation setups FF~\textup{(a)}, FH~\textup{(b)}, HF~\textup{(c)} and
HH~\textup{(d)}. Results are averaged over 30 simulated data sets for each
setup, where for each simulated data set we use the DIC-chosen
hmmSeq model. The proposed method controls the FDR closest and
slightly lower to the 45 degree line.}
\label{FDR}
\end{figure}

For each of the four competing methods, Figure~\ref{FDR} plots the
observed FDR against the nominal FDR. Ideally, we would like to
observe a 45 degree line through the origin in Figure~\ref{FDR} for
each method. Observed FDR of edgeR is substantially smaller than the
nominal FDR. The observed FDR for TSPM and baySeq, on the other hand,
are quite liberal: The FDR for TSPM always exceeds $40 \%$, while the
FDR for baySeq exceeds $35 \%$ for most values of nominal
FDR. Finally, FDR for hmmSeq is near and slightly lower than the 45
\,degree line. Therefore, hmmSeq is the method that performs best at
controling FDR.

{To investigate the robustness of hmmSeq to overdispersed data, we
simulated RNA-seq counts from a negative binomial distribution. This
distribution is assumed by both edgeR and baySeq. Assume that
$y\mid\lambda\sim \operatorname{Poisson}(\lambda)$ and $\lambda\mid r, \psi
\sim\operatorname{gamma}(r, (1-\psi)/\psi)$. Then,\vspace*{1pt} unconditionally, we
obtain $y\mid r, \psi\sim\operatorname{negative~binomial}(r, \psi)$. The\vspace*{1pt}
negative binomial mean is $m^{(1)}=r\psi/(1-\psi)$ and the variance
is $m^{(2)}=r\psi/(1-\psi)^2$. The variance $m^{(2)}=m^{(1)}(1 +
m^{(1)}/r)$ exceeds the mean $m^{(1)}$, reflecting overdispersion;
$\zeta=1/r$ is usually called the dispersion parameter.}

{The gene-specific magnitude factors $\bolds{\beta}$ and the
differential expression factors $\bolds{\Delta}$ were generated
from a hidden Markov model with parameters values given in
Tables~\ref{tabparametersbeta} and \ref{tabparametersdelta}. To
generate the negative binomial reads for each gene $i$, we first
generated the mean $m^{(1)}_{ijkc}=\lambda_{ijkc}$ by
equation~(\ref{lambda}) and the\vspace*{1pt} dispersion parameter $\zeta_i$ from
a gamma distribution [as in, e.g., \citet{kvam2012comparison}]. To
mimic the dispersion observed in real data sets, the shape parameter
and scale parameter of the gamma distribution were estimated by the
method of moments using the gene-wise dispersion estimates of
\citet{zeng2012divergent} (biological replicates) available from
edgeR. We computed the gamma distribution parameters $r=1/\zeta$
and $\psi=\zeta\mu/(1+\zeta\mu)$, and then hierarchically generated
negative binomial read counts for 12 chromosomes having 800 genes
each. This simulation procedure was replicated 30 times.}

%
\begin{figure}[t]

\includegraphics{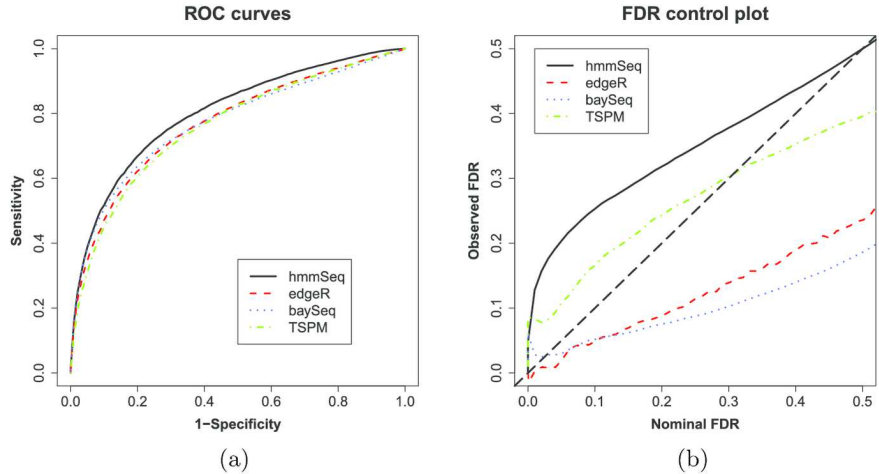}

\caption{For the negative binomial simulation study, the left panel
depicts the ROC curves for \mbox{DIC-}selected hmmSeq, edgeR, baySeq
and TSPM, and the right panel depicts the observed FDR versus nominal
FDR for DIC-selected hmmSeq, TSPM, edgeR and baySeq.
\textup{(a)}~ROC curves,
\textup{(b)}~FDR control plot.}\label{nb} \label{ROCnb} \label{FDRnb}
\end{figure}

{We fit four hmmSeq models to the negative binomial data. In
addition, we fit edgeR, baySeq and TSPM models to the data. DIC
chose the true model (HH in this case) 19 out of 30 times. The ROC
curves and FDR controls are plotted in Figure~\ref{nb}. In the FDR
control plot in Figure~\ref{FDRnb}(b), we find that none of the
methods are accurate. The hmmSeq FDR tends to be large for small
nominal FDR, converging to the 45 degree line as the nominal level
increases. In contrast, the observed FDR of baySeq and edgeR are
mostly lower than the nominal FDR. For the ROC plot in
Figure~\ref{ROCnb}(a), hmmSeq achieves the highest area under the ROC
curve than the competing methods, demonstrating its high reliability
in detecting DE genes.}

\section{Data analysis}

To illustrate the power and flexibility of our proposed RNA-seq
analysis method, we have applied the hmmSeq method to analyze three
data sets: \citet{marioni2008rna} (technical replicates),
\citet{zeng2012divergent} (biological replicates) and
\citet{henn2013high} (paired data). For each of these three data sets,
the treatment-specific replicate effects $\rho_{jk}$ are obtained by
the {upper-quartile} normalizing technique of
\citet{bullard2010evaluation}. In addition, we compare the results of
the hmmSeq analysis with results of TSPM [\citet{auer2011two}], baySeq
[\citet{hardcastle2009bayseq}] and edgeR [\citet
{robinson2010edger}] based
on their publicly available R package implementations.

\subsection{\texorpdfstring{\citet{marioni2008rna}}{Marioni et al. (2008)} data set}\label{Stechreplicate}

The \citet{marioni2008rna} data set contains RNA-seq data for five
technical replicates each of a single sample of kidney RNA (treatment
1) and liver RNA (treatment 2). Because genes with mostly small counts
are not informative about differential expression, we have applied the
filtering criterion of \citet{auer2011two} to eliminate the genes
whose total read counts were less than 10. Additionally, the Y
chromosome was ignored because many of its genes are transcribed on
other chromosomes and the genders of the subjects are unknown. This
yielded 17,076 genes for the analysis. Further, we applied the
quantile normalization of \citet{bullard2010evaluation} to preprocess
the data. We have fitted four hmmSeq models (FF, FH, HF and HH) to
the data. The DIC favors the FH model as the best, which implies
neighboring genes dependence with respect to the differential
expression parameter $\bolds{\Delta}$. Thus, in the remainder of
this section we present hmmSeq results based on the FH model.

%
\begin{figure}

\includegraphics{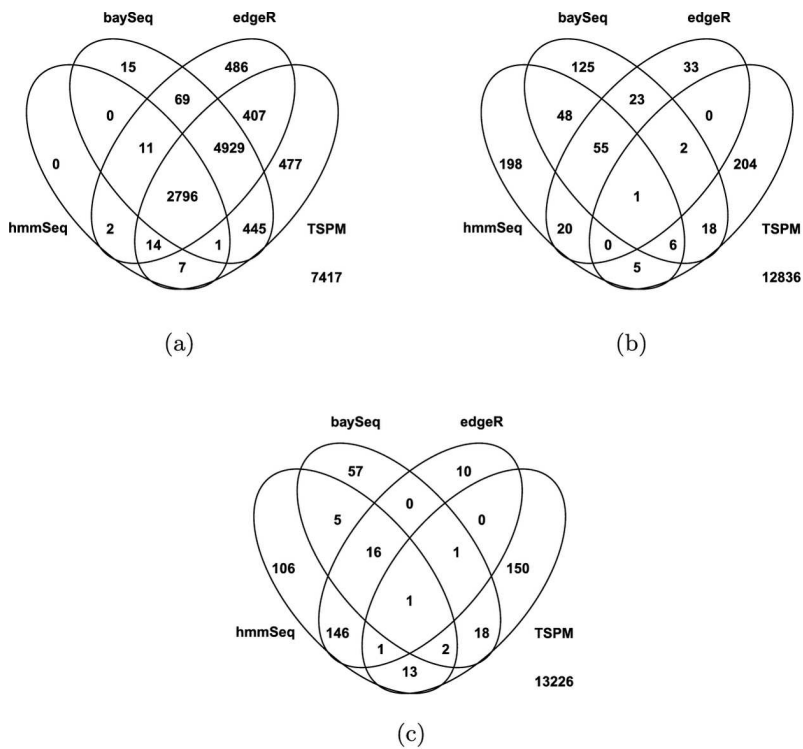}

\caption{Venn diagrams for the DE genes identified by the methods
hmmSeq, baySeq, edgeR and TSPM in the data analyses.
\textup{(a)}~Technical replicates of \citet{marioni2008rna},
\textup{(b)}~biological replicates of \citet{zeng2012divergent},
\textup{(c)}~paired biological replicates of \citet{henn2013high}.}\label{VANN} \label{VANNT} \label{VANNB} \label{VANNP}
\end{figure}

We have applied the hmmSeq, edgeR, baySeq and TSPM methods to the
\citet{marioni2008rna} data set with a nominal FDR of $q_0=0.001$
[threshold adopted by \citet{auer2011two}]. Recall that the
simulation study in Section~\ref{Ssimulation} had indicated that the
actual FDR of TSPM and baySeq is relatively insensitive to the value
of $q_0$ and is substantially higher when $q_0$ is small. The sets of
DE genes identified by the methods hmmSeq, edgeR, baySeq and TSPM are
summarized in Figure~\ref{VANNT}(a). The method TSPM detected 9076 DE
genes. In contrast, the hmmSeq method discovered only 2831 DE
genes.

A closer examination sheds light on the differing sets of genes
detected by TSPM and hmmSeq. Of the genes discovered by hmmSeq, as
many as 2818 genes (99.5\%) were also identified by the TSPM
method. Focusing on the 6258 genes identified as DE by TSPM but
not by hmmSeq, we find that TSPM flagged most of them as DE genes
because they have extreme values of mean $\log_2$-fold change (from
output of TSPM) for the treatments. In particular, we have observed
that for the $111$ genes with $\log_2$-fold change less than $-20$,
all five gene-specific reads for liver RNA were zero.
Table~\ref{0liver} lists 10 randomly selected genes from this set,
which reveals that the read counts for kidney, although positive, are
not in hundreds or thousands like typical DE genes. 
Thus, TSPM tends to classify genes with 0 observations under
any single condition as DE, while hmmSeq takes the variational
magnitude into consideration. The result for edgeR lies between
hmmSeq and TSPM.

Similarly, all $33$ genes with mean $\log_2$-fold change greater than
$20$ have zero counts for kidney and relatively small counts for
liver. The hmmSeq method called all of them non-DE, but TSPM classified
them as DE genes due to their high mean $\log_2$-fold changes. The
former call seems more reasonable, given that the read counts of truly
DE genes are typically several orders of magnitude higher.

%
\begin{sidewaystable}
\tablewidth=\textwidth
\tabcolsep=0pt
\caption{\citet{marioni2008rna} data set---ten randomly selected
genes with mean $\log_2$-fold change greater than 20 that were
identified as DE genes by the TSPM method. R$_j$L$_k$K denotes $j$th
run, $k$th replicate for kidney; R$_j$L$_k$L denotes $j$th run, $k$th
replicate for liver}\label{0liver}
\begin{tabular*}{\tablewidth}{@{\extracolsep{\fill}}lcccccccccc@{}}
\hline
\textbf{Gene ID} & $\mathbf{R_1L_1K}$& $\mathbf{R_1L_3K}$& $\mathbf{R_1L_7K}$& $\mathbf{R_2L_2K}$& $\mathbf{R_2L_6K}$ & $\mathbf{R_1L_2L}$& $\mathbf{R_1L_4L}$
       & $\mathbf{R_1L_6L}$& $\mathbf{R_1L_8L}$& $\mathbf{R_2L_3L}$\\
\hline
ENSG00000198693 & 2 & 1 & 2 & 2 & 8 & 0 & 0 & 0 & 0 & 0\\
ENSG00000162746 & 3 & 4 & 2 & 4 & 4 & 0 & 0 & 0 & 0 & 0\\
ENSG00000168243 & 4 & 4 & 1 & 3 & 3 & 0 & 0 & 0 & 0 & 0\\
ENSG00000188935 & 2 & 0 & 3 & 5 & 5 & 0 & 0 & 0 & 0 & 0\\
ENSG00000173284 & 5 & 2 & 5 & 2 & 3 & 0 & 0 & 0 & 0 & 0\\
ENSG00000114113 & 4 & 5 & 4 & 1 & 1 & 0 & 0 & 0 & 0 & 0\\
ENSG00000169836 & 5 & 2 & 3 & 2 & 3 & 0 & 0 & 0 & 0 & 0\\
ENSG00000170180 & 2 & 2 & 3 & 2 & 5 & 0 & 0 & 0 & 0 & 0\\
ENSG00000186952 & 3 & 4 & 2 & 4 & 2 & 0 & 0 & 0 & 0 & 0\\
ENSG00000164385 & 3 & 2 & 3 & 3 & 4 & 0 & 0 & 0 & 0 & 0\\
\hline
\end{tabular*}
\end{sidewaystable}
%

\subsection{\texorpdfstring{\citet{zeng2012divergent}}{Zeng et al. (2012)} data set}\label{Sbioreplicate}

We have applied hmmSeq to the data set of \citet{zeng2012divergent},
which contains samples from 2 regions of the human brain, frontal pole
and hippocampus, with 6 biological replicates for each region.
Data sets with biological replicates typically exhibit substantial
\textit{over-dispersion} or higher variability relative to the Poisson
distribution. Researchers often model extra-Poisson variability using
the binomial, negative binomial or Bayesian hierarchical Poisson
models. We account for over-dispersion through hierarchical priors on
the parameters in equation~(\ref{lambda}), for example, through random
differential expression
factors. 
There were 13,574 genes available for analysis after filtering (read
sums for all the libraries did not exceed 9). Further, we have
applied the quantile normalization of \citet
{bullard2010evaluation} to
preprocess the data. We have fitted four hmmSeq models (FF, FH, HF
and HH) to the \citet{zeng2012divergent} data set. DIC chooses HH as
the best model, which indicates neighboring genes dependence with
respect both to the differential expression parameter
$\bolds{\Delta}$ and to the expression level parameter
$\bolds{\beta}$. Thus, in the remainder of this section we
present hmmSeq results based on the HH model.

We have applied the hmmSeq, edgeR, baySeq and TSPM methods to the
data set of \citet{zeng2012divergent} with a nominal FDR $q_0=0.05$.
TSPM, edgeR and baySeq, respectively, identified $236$, $134$ and $278$
DE genes. The hmmSeq technique identified $333$ DE genes. The
overlapping set of DE genes for the methods are summarized in
Figure~\ref{VANNB}(b) and reveal a greater lack of agreement between the
methods than for the \citet{marioni2008rna} data set. Only~1 gene is
identified as DE by all four methods. This low level of agreement is
a result of the low overlap that TSPM has with the other methods. In
contrast, hmmSeq has relatively large overlap both with edgeR ($76$
genes) and baySeq ($110$ genes).

We have investigated the biological implications of the results
obtained with the hmmSeq analysis of differential expression of the
hippocampus to the frontal pole. Though we expect a modest amount of
differentially expressed genes, we do find some meaningful results
that are supported in the literature. There is an increase in gene
expression of Akt2 in the hippocampus compared to the frontal pole.
Akt2 is a gene involved in insulin signaling, which occurs in the
hippocampus [\citet{robertson2010insulin,agrawal2012metabolic}]. In
addition, Wnt7B is upregulated in the hippocampus where Wnt activity
has been implicated in signaling of hippocampal synapses
[\citet{gogolla2009wnt}]. Last, STAT5A is known to be expressed in
the hippocampus [\citet{kalita2013igf}], and our results show this
upregulation. Taken all together, the results our hmmSeq method
produced show biologically relevant genes when comparing the
hippocampus to the frontal pole. In addition, we have conducted a
functional analysis using DAVID [\citeauthor{da2009bioinformatics} (\citeyear{da2009bioinformatics,da2009systematic})]; both our hmmSeq method and
edgeR identified
differentially expressed genes that are known to be expressed in brain
tissue. 
The biological experiment presented tries to identify genes that are
differentially expressed between two types of brain tissue. These
results, taken in perspective with the biological experiment, suggest
that the genes identified as differentially expressed via these two
methods are relevant to the biological problem and help support the
validity and accuracy of our predictions.

\subsection{\texorpdfstring{\citet{henn2013high}}{Henn et al. (2013)} paired data set}
\label{Spairedreplicate}

Here we illustrate the application of the hmmSeq method to paired
data sets with an analysis of a subset of an RNA-seq data set obtained
by \citet{henn2013high} on immune response to a trivalent influenza
vaccine. The original data set contains RNA-seq data from B cell
samples for five subjects before vaccination (day 0) and for each of
10 days after vaccination (days 1 through 10). We consider a paired
subset of three previously vaccinated subjects in the original data set,
where we compare gene expression before vaccination to gene expression
after vaccination. Since peak B cell response usually appears \mbox{5--9}
days post-vaccination, we apply our hmmSeq method to identify B cell
gene differential expression between day 0 and day 7.



For the hmmSeq analysis, we first estimate the variance
$\sigma^2_{\varepsilon}$ of subject-specific random effects from the
data. Specifically, for each gene we have fitted a generalized linear
mixed model with random subject effects, resulting in an estimated
random effects variance for each gene. We then use the median of these
estimates of random effects variances as an empirical Bayes estimate
of ${\sigma}^2_{\varepsilon}$. In addition, we have fitted the four
hmmSeq models FF, FH, HF and HH. We have found that the DIC favors the
FF model, that is, a finite mixture model without neighboring genes
dependence. Thus, in the remainder of this section we present hmmSeq
results based on the FF model.


We have analyzed this immune response data set using a nominal FDR of
$0.05$. To accommodate the paired data structure, in the edgeR
analysis we include subject-specific fixed effects. Such edgeR
analysis identifies $175$ genes as differentially expressed. The TSPM
that we used ignores the paired structure and treats all observations
for each gene as independent, which identifies a total of $186$ genes.
Finally, in a paired baySeq analysis $100$ genes are flagged to be
DE. Figure~\ref{VANNP}(c) presents a Venn diagram that summarizes the
results for TSPM, edgeR, baySeq and hmmSeq.

%
\begin{table}[b]
\tabcolsep=0pt
\caption{\citet{henn2013high} data set---overlap of plasma cell gene
signature (PCgs) set with genes identified by hmmSeq, edgeR, baySeq
and TSPM}\label{taboverlap}
\begin{tabular*}{\tablewidth}{@{\extracolsep{\fill}}@{}lcccc@{}}
\hline
& \textbf{hmmSeq} & \textbf{edgeR} & \textbf{baySeq} & \textbf{TSPM} \\
\hline
PCgs & 130 & 121 & 7 & 1\\
\hline
\end{tabular*}
\end{table}

In order to further evaluate the competing methods, we compare their
results to those found by \citet{henn2013high}. \citet{henn2013high}
used the RNA-seq data set from all 11 days, whereas we used only the
data from days 0 and 7. Thus, here we use the results of
\citet{henn2013high} as a benchmark. Specifically, \citet
{henn2013high}
identified a set of $742$ genes as what they call the plasma cell gene
signature (PCgs), that is, genes that have a common significant
time-varying signature. Hence, in Table~\ref{taboverlap} we list the
overlap of the PCgs set with the genes identified as differentially
expressed by hmmSeq, edgeR, baySeq and TSPM. Our proposed hmmSeq
method obtains the largest overlap with PCgs set ($130$ genes), and
edgeR overlaps $121$ genes with PCgs. We recall from
Section~\ref{Ssimulation} that hmmSeq and edgeR are the two methods
with the highest area under the ROC curve. Thus, the overlap with the
PCgs set shows the power of DE genes identification of the proposed
hmmSeq method.

\section{Conclusion}

We propose hmmSeq, a method based on Bayesian hierarchical models for
detecting DE genes between two treatments for paired or nonpaired data
in RNA-seq analyses. The approach employs hidden Markov models to
account for the statistical dependence between the gene counts of
neighboring genes observed in many RNA-seq data sets. The hmmSeq model
can be applied to studies with either biological or technical
replicates, automatically adjusting for any overdispersion relative to
the Poisson distribution. Through simulated and real data sets, we
compare and contrast the performance of hmmSeq with some well-known
methods in the literature, demonstrating the reliability and success
of our approach in the identification of DE genes.

We have developed DIC-based model selection to decide for each data set
whether HMM or FMM should be used to model gene expression magnitude
and/or differential expression. For the \citet{marioni2008rna}
data set
the DIC-selected model is the FH model, for the
\citet{zeng2012divergent} data set the DIC-selected model is the HH
model, and for the \citet{henn2013high} data set the DIC-selected
model is
the FF model. Thus, for one data set there appears to be neighboring
genes dependence in expression magnitude. Even more important, for two
data sets there is evidence of neighboring genes dependence in
differential
expression. 
This co-differential expression can be justified by how ancient
species organized their genomes and by evolution. Specifically, more
ancient species, such as bacteria, organize their genomes based on
operons, where genes involved in the same process or needed at the same
time are transcribed in tandem
[\citet{alberts1994molecular}]. Throughout evolution, operons have been
divided into individual genes, but genes involved in the same process
now reside in gene clusters [\citet{hurst2004evolutionary}]. Thus,
neighboring genes tend to be jointly differentially
expressed. 

To further examine spatial genomic dependence (and clustering) among
detected DE genes, we have devised the following statistical
test. Consider any detected DE gene and the next detected DE gene in
the chromosome as neighboring DE genes. Consider the distance between
two neighboring DE genes as the number of non-DE genes between them. If
there is no spatial dependence, then all the distances between any two
neighboring DE genes should be a random sample from a geometric
distribution. Hence, to test for spatial dependence, we collect all the
distances between neighboring DE genes and conduct a goodness-of-fit
test of the hypothesis that the empirical distribution equals the null
theoretical geometric distribution. We use this procedure to test for
spatial genomic dependence for DE calls from edgeR and hmmSeq. First,
we performed this test for the \citet{henn2013high} data set for which
hmmSeq prefers finite mixture model and spatial independence. The
spatial genomic dependence test for DE calls from edgeR and hmmSeq
yields $p$-values equal to 0.2201 and 0.5178, respectively, further
supporting hmmSeq suggestion of spatial independence. Second, we
performed this test for the two real data sets for which hmmSeq prefers
spatial dependence, that is, the \citet{marioni2008rna} and the
\citet{zeng2012divergent} data sets. For the \citet{marioni2008rna}
data set, the spatial dependence test for DE calls from both hmmSeq and
edgeR yield $p$-values smaller than 2.2e--16. That is, even though
edgeR does not account for spatial dependence, its detected DE genes
for the \citet{marioni2008rna} data set cluster spatially. For the
\citet{zeng2012divergent} data set, edgeR only detected 134 DE genes
which did not provide enough power for the goodness-of-fit data set
($p$-value greater than 0.9). In contrast, there is strong
statistical evidence that the 333 genes identified by hmmSeq as DE
cluster spatially ($p$-value smaller than 2.2e--16). Therefore, these
data sets point to the need to consider spatial genomic dependence in
studies of differential gene expression.

In addition to genomic spatial dependence among genes based on genomic
position, for future research work we plan to extend hmmSeq to include
other sources of dependence among genes. Recent experimental
techniques such as HiC and ChIA-PET allow for the identification of
explicit promoter--promoter and promoter--enhancer--promoter interactions
[\citet{pmid22365496,pmid23817049,pmid25160912}]. In addition, we note
that genes that belong to the same active functional pathways tend to
be co-expressed [\citet{pmid23284898}]. This extension of hmmSeq may
need a non-Markovian spatial correlation model. We leave this
challenging inferential problem to future research.

DE gene call lists are frequently used in downstream pathway function
calls in what is known as functional enrichment analysis. Because
functional enrichment analysis methods usually assume independence of
DE gene calls, caution needs to be taken when using the DE gene call
lists generated by hmmSeq. When hmmSeq decides that the best model is
a finite mixture model without spatial dependence, then one can use
hmmSeq's DE gene call list without any concern. However, when hmmSeq
decides that a spatial dependence model is warranted, then the
assumption of independence no longer holds. This opens up a tremendous
opportunity for future research that performs joint differential
expression gene calls and functional enrichment analysis. We envision
this joint analysis may be implemented by extending hmmSeq to
incorporate information on functional pathway networks.

In terms of computational time, on a desktop with a 2.3~GHz processor and
4~GB memory, hmmSeq takes approximately 3 hours to analyze a 1200-gene
chromosome. Although it does take a longer time than other methods,
hmmSeq often achieves a higher accuracy of DE gene detection than
other methods by a realistic model that allows for spatial genomic
dependence. Moreover, the computational time of all the considered
statistical methods is negligible when compared to the time (on the
order of months or years) required by subject-matter scientists to
perform experiments to obtain RNA-seq data. Furthermore, to limit the
computational time the hmmSeq analysis can be performed in parallel
for individual chromosomes. Finally, when compared to the high costs
of RNA-seq extraction, the information gains obtained by the hmmSeq
methodology seem well worth the relatively low computational costs.

The hmmSeq method we propose relies on a single user-specified
``tuning'' parameter $q_0$, that is, the nominal false discovery rate. A
default value for $q_0$ between 0.001 or 0.05 has produced
satisfactory results for all the data sets, real or simulated, that we
have analyzed, facilitating ``black box'' applications of
hmmSeq. Future work will focus on extending hmmSeq to investigations
with more than two treatments. {An R package implementing the hmmSeq
framework will be submitted to CRAN upon publication of the
manuscript}.


\section*{Acknowledgments}
The authors would like to thank
Martin Zand and Stephen Welle at the University of
Rochester Medical Center for providing the vaccine
data in Section \ref{Spairedreplicate}.

\begin{supplement}[id=suppA]
\stitle{Supplement to ``hmmSeq: A hidden Markov model for detecting differentially expressed genes from RNA-seq data.''\\}
\slink[doi]{10.1214/15-AOAS815SUPP} 
\sdatatype{.zip}
\sfilename{aoas815\_supp.zip}
\sdescription{The R code for hmmSeq.}
\end{supplement}

%

\printaddresses
\end{document}